# Thermodynamic properties of ideal Bose gas trapped in different external power-law potentials under generalized uncertainty principle*

Ya-Ting Wang, He-Ling Li[†]

(School of Physics and Electronic-Electrical Engineering, Ningxia University, Yinchuan 750021, China)

Abstract

Significant evidence is available to support the quantum effects of gravity that leads to the generalized uncertainty principle (GUP) and the minimum observable length. Usually quantum mechanics, statistical physics doesn't take gravity into account. Thermodynamic properties of ideal Bose gases in different external power-law potentials are studied under the GUP with statistical physical method. Critical temperature, internal energy, heat capacity, entropy, particles number of ground state and excited state are calculated analytically to ideal Bose gases in the external potentials under the GUP. Below the critical temperature, taking the rubidium and sodium atoms ideal Bose gases whose particle densities are under standard and experimental conditions, respectively, as examples, the relations of internal energy, heat capacity and entropy with temperature are analyzed numerically. Theoretical and numerical calculations show that: 1) the GUP leads to an increase in the critical temperature. 2）When the temperature is lower than the critical temperature and slightly higher than 0K, the GUP's amendments to internal energy, heat capacity and entropy etc. are positive. As the temperature increases to a certain value, these amendments become negative. 3)The external potentials can increase or decrease the influence of the GUP on thermodynamic properties. When $\varepsilon$=1J, $\varepsilon$ is the quantity that reflects the external potential intensity, and atomic density $n = 2.687 \times 10^{25} \mathrm{m}^{-3}$, the GUP's amendments to the internal energy, heat capacity and entropy of the rubidium atoms ideal Bose gas first decrease and then increase with the increase of $X$ (where $X \equiv \Sigma_i 1/t_i$ is sum of the reciprocal of the exponents of the power function). In 3-dimensional harmonic potential, the relative correction term of the GUP is 26 orders of magnitude larger than that of a free-particle system in a fixed container. 4）When $\varepsilon \approx 10^{-31}$J and $n \approx 10^{20}$m$^{-3}$ (which are the experimental data when BEC was first verified by sodium atomic gas), the influence of the GUP can be completely ignored. 5) Under certain conditions, the GUP may become the dominant factor governing the thermodynamic properties of the system.

PACS：05.30.–d, 03.75.Hh

* Project supported by the Major Innovation Projects for Building First-class Universities in China's Western Region (Grant No. ZKZD2017006); Natural science foundation project of ningxia education department (Grant No. NGY0017054).

† Corresponding author E-mail: ningxiayclhl@163.com   Telephone: 13995100981

## 1. Introduction

Matter exhibits superfluid and superconducting characteristics at appropriately low temperature. Theories and experiments have shown that superfluidity and superconductivity are closely related to the "cooperation" between quantum particles, such as the well-known conventional superconductor microscopic theory (BCS theory, the name comes from Bardeen, Cooper and Schrieffer) and Bose-Einstein condensation theory (BEC theory). The experimental verification of the BEC [1-3] has improved our understanding of the nature of macro-quantum objects [4-9].

Applying an external potential is important for restricting and studying quantum gases [10, 11]. By changing the shape and intensity of the external potential, we can artificially control the scattering length between atoms, which is to control the interaction between atoms [12], and regulate various behaviors of quantum systems. Many scholars have studied the thermodynamic properties of ideal boson gas and weakly interacting boson gas under different limiting potentials [13-18].

When studying the "cooperation" between particles in the context of statistical physics and quantum mechanics, because the influence of gravity is minor, we usually ignore gravitational effects. However, gravity is everywhere. In a quantum theory of gravity, the Heisenberg algebra of quantum mechanics is replaced by a deformed Heisenberg algebra, and fundamental commutation is modified so that the uncertainty principle in regular quantum mechanics is modified to the generalized uncertainty principle (GUP) [19]. There exists a minimal observed length, which is a basic quantity closely related to the structure of space-time [20]. Considerable evidence (from string theory [21], loop quantum gravity [22], and non-commutative geometry [23]) is available to confirm the GUP.

When considering the quantum effect of gravity, the density of quantum states in statistical physics is modified, which may have a significant influence on theoretical reasoning regarding the thermodynamic properties of the system. Some studies in the literature have applied the GUP to systems of blackbody radiation [24], harmonic oscillators [25-26], ideal gases [24, 27], and astrophysics [28-30]. And some theoretical results have been subversive to traditional concepts, for example, heat capacity tends to zero at extremely high temperatures [24]. However, the heat capacity calculated by traditional statistical physics is the result of the energy equality theorem at normal high-temperature conditions, that is, it is a temperature-independent constant. For a classic ideal gas, the number of particles in an excited state of the system tends to a certain value as the temperature tends to infinity. That is, if the number of particles in the system increases, they "condense" in the ground state (the condensation as temperature tends to infinity!). The energy of the system no longer increases at a limit temperature, and an upper energy limit appears [27]. The above results are completely different from those deduced by traditional statistical physics, thus subverting it.

In this paper, the thermodynamic properties of ideal Bose gases in different external power-law potentials are studied under the GUP. Thermodynamic quantities, such as the mean particle number in ground state and excited state, critical temperature, internal energy, heat capacity and entropy, are calculated analytically and analyzed numerically. The influences of different external power-law potentials and of the GUP on the thermodynamic properties of Bose systems are discussed.

## 2 Critical temperature of ideal Bose gases in different external power-law potentials under the GUP

Consider a system which consists of an ideal Bose gas trapped in an external potential. Single-particle energy in the external potential may be expressed as

$$\varepsilon(\boldsymbol{p},\boldsymbol{r}) = \frac{\boldsymbol{p}^2}{2m} + V(\boldsymbol{r}), \tag{1}$$

where $\boldsymbol{p}=(p_x, p_y, p_z)$ and $\boldsymbol{r}=(x,y,z)$ are the momentum and the coordinates of particles in three-dimensional space, respectively, $m$ is the rest mass of a particle, $\boldsymbol{p}^2 = p_x^2 + p_y^2 + p_z^2$. The external power-law potentials may be expressed as

$$V(\boldsymbol{r}) = \varepsilon_1 |x/L_1|^{t_1} + \varepsilon_2 |y/L_2|^{t_2} + \varepsilon_3 |z/L_3|^{t_3}. \tag{2}$$

Where $t_i$, $\varepsilon_i$, and $L_i$ ($i=1,2,3$) are positive constants describing the external potentials. When the number of particles is large, and potential energy of the particles is smaller than the kinetic energy of particles in the external potential (this condition is often met in relevant experiment [1]), and we can use the Thomas-Fermi semi-classical approximation. In this instance, the logarithm of the grand partition function may be

expressed as

$$\ln \Xi = -\frac{g}{h^3} \int \ln[1 - e^{-\alpha-\beta\varepsilon}] d\mathbf{p} d\mathbf{r}, \qquad (3)$$

where $h$ is Planck constant, $g$ is spin degeneracy, $\beta = 1/k_B T$, $\alpha = -\beta\mu$, $k_B$ is Boltzmann constant, $\mu$ is chemical potential, and $T$ is absolute temperature. $d\mathbf{p} = dp_x dp_y dp_z$, $d\mathbf{r} = dxdydz$.

From Ref. [19], one can easily derive the simplest form of the GUP with modified commutation

$$\Delta x \Delta p \geq \frac{\hbar}{2}(1 + A(\Delta p)^2), \qquad (4)$$

Where $\hbar = h/2\pi$. Equation (4) implies a minimum observable length [19,24], and the relationship between the positive parameter $A$ and the minimum observable length $\Delta x_{min}$ is expanded as

$$\Delta x_{min} = \hbar\sqrt{A}. \qquad (5)$$

Where $\Delta x_{min}$ is closely related to space-time structure [20].

When considering the GUP, the density of states should include a correction factor [24-27, 31-32]: $(1 + Ap^2)^{-3}$. Then the logarithm of the grand partition function of the boson system is corrected as

$$\ln \Xi = -\frac{g}{h^3} \int \frac{\ln(1 - e^{-\beta(\varepsilon-\mu)})}{(1 + Ap^2)^3} d\mathbf{p}d\mathbf{r}. \qquad (6)$$

Because in many cases the effect of the GUP is small, the amendment factor $(1 + Ap^2)^{-3}$ can be expanded by momentum [27,32].

$$(1 + Ap^2)^{-3} = 1 - 3Ap^2 + \frac{3(3+1)}{2!}(Ap^2)^2 - \frac{3(3+1)(3+2)}{3!}(Ap^2)^3 + \cdots. \qquad (7)$$

Substituting equation (7) into equation (6), using spherical coordinate and integrating, we can obtain

$$\ln \Xi = \frac{gV^*}{\lambda^3}[g_{\eta+1}(z_r) - 9g_{\eta+2}(z_r)mAk_B T + 90g_{\eta+3}(z_r)(mAk_B T)^2 - \cdots]. \qquad (8)$$

Where $\lambda = h/\sqrt{2\pi mk_B T}$ is the thermal wavelength, $z_r = e^{\beta\mu}$ is fugacity.

$$V^* = \prod_{i=1}^{3} \frac{2L_i}{(\beta\varepsilon_i)^{1/t_i}} \Gamma(\frac{1}{t_i} + 1), \qquad (9)$$

$$\eta = X + \frac{3}{2}, \quad X = \sum_{i=1}^{3} \frac{1}{t_i}, \tag{10}$$

$g_l(z)$ is the Bose integration

$$g_l(z) = \frac{1}{\Gamma(l)} \int_0^\infty \frac{t^{l-1}}{z^{-1}e^t - 1} dt \xrightarrow{z \leq 1} \sum_{j=1}^\infty \frac{z^j}{j^l}, \tag{11}$$

$g_l(z)$ meets the following conditions

$$g_l(z) = z \frac{dg_{l+1}(z)}{dz}, \quad (l > 1), \tag{12}$$

$\Gamma(l)$ is a gamma function

$$\Gamma(l) = \int_0^\infty y^{l-1} e^{-y} dy. \tag{13}$$

When the GUP is not considered ($A=0$), the results obtained by equation (7) and thermodynamic formulas are consistent with those in Refs. [13,16,17].

Using a basic formula of statistical mechanics, the average number of particles of the system can be expressed as

$$N = -\frac{\partial \ln \Xi}{\partial \alpha} = \frac{gV^*}{\lambda^3}[g_\eta(z_r) - 9g_{\eta+1}(z_r)mAk_BT + 90g_{\eta+2}(z_r)(mAk_BT)^2 - \cdots]. \tag{14}$$

When the number of particles $N$ is constant, the chemical potential $\mu$ increases with a decrease in temperature. When temperature $T$ decreases to $T_c$, $\mu \to 0$, $z_r \to 1$, and Bose–Einstein condensation (BEC) begins to occur. At this time, the number of particles $N_0$ of the ground state with zero energy and momentum can still be ignored. Equation (14) may be expressed as

$$N = \frac{gV^*}{\lambda_c^3}[\zeta(\eta) - 9\zeta(\eta+1)mAk_BT_c + 90\zeta(\eta+2)(mAk_BT_c)^2 - \cdots]. \tag{15}$$

Where $g_l(1) = \sum_{j=1}^\infty 1/j^l = \zeta(l)$ is the Riemann-zeta function that converges when $l > 1$ and diverges when $l \leq 1$, $\lambda_c = h/\sqrt{2\pi mk_BT_c}$. We let

$$n^* = \frac{N}{V'}; \quad V' = \prod_{i=1}^{3} \frac{2L_i}{(\varepsilon_i)^{1/t_i}} \Gamma(\frac{1}{t_i} + 1). \tag{16}$$

When the GUP is not considered ($A=0$), the BEC temperatures of ideal Bose gases in external power-law potentials in 3D space can be obtained as

$$T_{c0} = \frac{1}{k_B}\left(\frac{n^* h^3}{g(2\pi m)^{3/2}\zeta(\eta)}\right)^{\frac{1}{\eta}}$$

$$= \frac{1}{k_B}\left[\frac{Nh^3}{\zeta(\eta)(2\pi m)^{3/2}}\prod_{i=1}^{3}\frac{\varepsilon_i^{1/t_i}}{g(2L_i)\Gamma(1/t_i+1)}\right]^{1/\eta}. \quad (17)$$

By applying the iterative method to equation (15), when considering the GUP and retaining only its second-order amendment, the BEC temperature $T_c$ of the ideal Bose gases in external power-law potentials can be expressed as

$$T_c = T_{c0}\left\{1 + \frac{9}{\eta}\frac{\zeta(\eta+1)}{\zeta(\eta)}mAk_BT_{c0} + \frac{9}{\eta}\left[\left(\frac{9}{2\eta}+\frac{9}{2}\right)\frac{\zeta^2(\eta+1)}{\zeta^2(\eta)} - \frac{10\zeta(\eta+2)}{\zeta(\eta)}\right](mAk_BT_{c0})^2 + \cdots\right\}. \quad (18)$$

The corrections to the BEC temperatures of the ideal Bose gases in external power-law potentials by the GUP can be expressed as

$$\Delta T_c = T_{c0}\left\{\frac{9}{\eta}\frac{\zeta(\eta+1)}{\zeta(\eta)}mAk_BT_{c0} + \frac{9}{\eta}\left[\left(\frac{9}{2\eta}+\frac{9}{2}\right)\frac{\zeta^2(\eta+1)}{\zeta^2(\eta)} - \frac{10\zeta(\eta+2)}{\zeta(\eta)}\right](mAk_BT_{c0})^2 + \cdots\right\}. \quad (19)$$

## 3 Thermodynamic properties of ideal Bose gases trapped in different external power-law potentials under the GUP

### 3.1 The thermodynamic functions at $T > T_c$

Using the grand partition function of equation (8) and the particle number of equation (14), the internal energy $U$ of the system can be obtained

$$U = kT^2\frac{\partial \ln\Xi}{\partial T}$$

$$= \eta Nk_BT\frac{g_{\eta+1}(z_r)}{g_\eta(z_r)}\left\{1 + 9\left[\frac{g_{\eta+1}(z_r)}{g_\eta(z_r)} - \frac{\eta+1}{\eta}\frac{g_{\eta+2}(z_r)}{g_{\eta+1}(z_r)}\right]mAk_BT\right.$$

$$\left. + \left[81\frac{g_{\eta+1}^2(z_r)}{g_\eta^2(z_r)} - \left(90 + 81\frac{\eta+1}{\eta}\right)\frac{g_{\eta+2}(z_r)}{g_\eta(z_r)} + \frac{90(\eta+2)}{\eta}\frac{g_{\eta+3}(z_r)}{g_{\eta+1}(z_r)}\right](mAk_BT)^2 + \cdots\right\}. \quad (20)$$

Using equation (20), the heat capacity $C$ can be obtained

$$C = \frac{\partial U}{\partial T}$$

$$= \eta N k_B \left\{ (\eta+1)\frac{g_{\eta+1}(z_r)}{g_\eta(z_r)} - \eta \frac{g_\eta(z_r)}{g_{\eta-1}(z_r)} + 9\left[(\eta+2)\frac{g_{\eta+1}(z_r)}{g_{\eta-1}(z_r)} + (\eta+1)\frac{g_{\eta+1}^2(z_r)}{g_\eta^2(z_r)} - \eta \frac{g_\eta^2(z_r)}{g_{\eta-1}^2(z_r)} \right. \right.$$

$$\left. - \frac{(\eta+1)(\eta+2)}{\eta}\frac{g_{\eta+2}(z_r)}{g_\eta(z_r)} \right] mAk_B T + 9\left[(19\eta+18)\frac{g_{\eta+1}(z_r)g_\eta(z_r)}{g_{\eta-1}^2(z_r)} + \frac{10(\eta+2)(\eta+6)}{\eta}\frac{g_{\eta+3}(z_r)}{g_\eta(z_r)} \right.$$

$$\left. -10(\eta+4)\frac{g_{\eta+2}(z_r)}{g_{\eta-1}(z_r)} - 9\eta \frac{g_\eta^3(z_r)}{g_{\eta-1}^3(z_r)} + 9(\eta+1)\frac{g_{\eta+1}^3(z_r)}{g_\eta^3(z_r)} - \frac{(\eta+1)(19\eta+18)}{\eta}\frac{g_{\eta+2}(z_r)g_{\eta+1}(z_r)}{g_\eta^2(z_r)} \right.$$

$$\left. \left. + \frac{(16\eta^2-9)}{\eta}\frac{g_{\eta+1}^2(z_r)}{g_{\eta-1}(z_r)g_\eta(z_r)} \right](mAk_B T)^2 + \cdots \right\}. \tag{21}$$

The entropy $S$ may be expressed as

$$S = k_B (\ln \Xi + \alpha N + \beta U)$$

$$= N k_B \frac{g_{\eta+1}(z_r)}{g_\eta(z_r)} \left\{ (\eta+1) + 9\left[(\eta+1)\frac{g_{\eta+1}(z_r)}{g_\eta(z_r)} - (\eta+2)\frac{g_{\eta+2}(z_r)}{g_{\eta+1}(z_r)} \right] mAk_B T \right.$$

$$\left. + 9\left[9(1+\eta)\frac{g_{\eta+1}^2(z_r)}{g_\eta^2(z_r)} - (19\eta+28)\frac{g_{\eta+2}(z_r)}{g_\eta(z_r)} + 10\frac{(\eta+3)g_{\eta+3}(z_r)}{g_{\eta+1}(z_r)} \right](mAk_B T)^2 - \ln z_r + \cdots \right\}. \tag{22}$$

The above calculations are obtained by ignoring the number of particles $N_0$ in the ground state. So this is the thermodynamic property at $T > T_c$. When the temperature is very low, that is, when $T \leq T_c$, $N_0$ increases with the decrease of temperature. $N_0$ cannot be ignored and the BEC occurs.

### 3.2 The thermodynamic functions at $T < T_c$

When $T < T_c$, $\mu \approx 0$, $z_r \approx 1$, the particles number $N_e$ in excited state is

$$N_e = \frac{gV^*}{\lambda^3}\left[\zeta(\eta) - 9\zeta(\eta+1)mAk_B T + 90\zeta(\eta+2)(mAk_B T)^2 - \cdots \right]. \tag{23}$$

$N_e$ can be expressed by the total average number of particles $N$

$$N_e = N\left(\frac{T}{T_c}\right)^\eta \left[ 1 + \frac{9\zeta(\eta+1)mAk_B}{\zeta(\eta)}(T_c - T) + \frac{90\zeta(\eta+2)(mAk_B)^2}{\zeta(\eta)}(T^2 - T_c^2) \right.$$

$$\left. + \left(\frac{9\zeta(\eta+1)mAk_B}{\zeta(\eta)}\right)^2 (T_c^2 - TT_c) + \cdots \right]. \tag{24}$$

That is, $N_e$ decreases with the decrease of temperature. The number of particles $N_0$ in the ground state can be expressed as

$$N_0 = N - N_e = N\left\{1 - \left(\frac{T}{T_c}\right)^{\eta}\left[\left(1 + \frac{9\zeta(\eta+1)mAk_B}{\zeta(\eta)}\right)(T_c - T)\right.\right.$$
$$\left.\left. + \frac{90\zeta(\eta+2)(mAk_B)^2}{\zeta(\eta)}(T^2 - T_c^2) + \left(\frac{9\zeta(\eta+1)mAk_B}{\zeta(\eta)}\right)^2(T_c^2 - TT_c) + \cdots\right]\right\}. \quad (25)$$

From equation (25), we get

$$\frac{N_0}{N} = 1 - \left(\frac{T}{T_c}\right)^{\eta}\left[\left(1 + \frac{9\zeta(\eta+1)mAk_B}{\zeta(\eta)}\right)(T_c - T) + \frac{90\zeta(\eta+2)(mAk_B)^2}{\zeta(\eta)}(T^2 - T_c^2)\right.$$
$$\left. + \left(\frac{9\zeta(\eta+1)mAk_B}{\zeta(\eta)}\right)^2(T_c^2 - TT_c) + \cdots\right] \quad (26)$$

When $T < T_c$, only the particles in the excited state contribute to the internal energy. The internal energy is

$$U = \eta Nk_B T\left(\frac{T}{T_c}\right)^{\eta}\frac{\zeta(\eta+1)}{\zeta(\eta)}\left\{1 + 9\left[\frac{\zeta(\eta+1)}{\zeta(\eta)} - \frac{\eta+1}{\eta}\frac{\zeta(\eta+2)}{\zeta(\eta+1)}\right]mAk_B T\right.$$
$$+ 9\frac{\zeta(\eta+1)}{\zeta(\eta)}mAk_B(T_c - T) + \left(9\frac{\zeta(\eta+1)}{\zeta(\eta)}\right)^2(mAk_B)^2(T_c^2 - T_cT) + \frac{90\zeta(\eta+2)}{\zeta(\eta)}(mAk_B)^2$$
$$\times(T^2 - T_c^2) + 81\frac{\zeta(\eta+1)}{\zeta(\eta)}\left[\frac{\zeta(\eta+1)}{\zeta(\eta)} - \frac{\eta+1}{\eta}\frac{\zeta(\eta+2)}{\zeta(\eta+1)}\right](mAk_B)^2(T_cT - T^2)$$
$$\left. + 9\left[9(1+\eta)\frac{\zeta^2(\eta+1)}{\zeta^2(\eta)} - (19\eta+28)\frac{\zeta(\eta+2)}{\zeta(\eta)} + 10(\eta+3)\frac{\zeta(\eta+3)}{\zeta(\eta+1)}\right](mAk_B T)^2 + \cdots\right\}. \quad (27)$$

When $T < T_c$, the heat capacity $C$ and the entropy $S$ are, respectively

$$C = \eta Nk_B\left(\frac{T}{T_c}\right)^{\eta}\frac{\zeta(\eta+1)}{\zeta(\eta)}\left\{(\eta+1) + 9(\eta+1)\frac{\zeta(\eta+1)}{\zeta(\eta)}mAk_B T_c\right.$$
$$- 9(\eta+2)\frac{\eta+1}{\eta}\frac{\zeta(\eta+2)}{\zeta(\eta+1)}mAk_B T + \left(9\frac{\zeta(\eta+1)}{\zeta(\eta)}\right)^2(mAk_B)^2\left[(\eta+1)T_c^2 - (\eta+2)TT_c\right]$$
$$+ 9\frac{\zeta(\eta+1)}{\zeta(\eta)}\left[9\frac{\zeta(\eta+1)}{\zeta(\eta)} - \frac{\eta+1}{\eta}\frac{\zeta(\eta+2)}{\zeta(\eta+1)}\right](mAk_B)^2\left[(\eta+2)TT_c - (\eta+3)T^2\right]$$
$$+ 90\frac{\zeta(\eta+2)}{\zeta(\eta)}(mAk_B)^2\left[(\eta+3)T^2 - (\eta+1)T_c^2\right] + 9(\eta+3)\left[9(1+\eta)\frac{\zeta^2(\eta+1)}{\zeta^2(\eta)}\right.$$
$$\left.\left. - (19\eta+28)\frac{\zeta(\eta+2)}{\zeta(\eta)} + 10(\eta+3)\frac{\zeta(\eta+3)}{\zeta(\eta+1)}\right](mAk_B T)^2 + \cdots\right\}. \quad (28)$$

$$S = Nk_B \left(\frac{T}{T_c}\right)^\eta \frac{\zeta(\eta+1)}{\zeta(\eta)} \left\{ (\eta+1) + 9 \left[ (\eta+1)\frac{\zeta(\eta+1)}{\zeta(\eta)} - (\eta+2)\frac{\zeta(\eta+2)}{\zeta(\eta+1)} \right] mAk_B T \right.$$

$$+ (\eta+1)\frac{9\zeta(\eta+1)}{\zeta(\eta)} mAk_B (T_c - T) + 9 \left[ 9(1+\eta)\frac{\zeta^2(\eta+1)}{\zeta^2(\eta)} - (19\eta+28)\frac{\zeta(\eta+2)}{\zeta(\eta)} \right.$$

$$\left. + 10(\eta+3)\frac{\zeta(\eta+3)}{\zeta(\eta+1)} \right] (mAk_B T)^2 + 90(\eta+1)\frac{\zeta(\eta+2)}{\zeta(\eta)} (mAk_B)^2 (T^2 - T_c^2)$$

$$+ \frac{81\zeta(\eta+1)}{\zeta(\eta)} \left[ (\eta+1)\frac{\zeta(\eta+1)}{\zeta(\eta)} - (\eta+2)\frac{\zeta(\eta+2)}{\zeta(\eta+1)} \right] (mAk_B)^2 (T_c T - T^2)$$

$$\left. + (\eta+1) \left( \frac{9\zeta(\eta+1)}{\zeta(\eta)} \right)^2 (mAk_B)^2 (T_c^2 - TT_c) + \cdots \right\}. \qquad (29)$$

## 4. Results and discussion

From equations (18)-(29), it can be seen that the corrections of the critical temperature and thermodynamic properties of ideal Bose gases trapped in different external power-law potentials by the GUP are reflected by the series of $mAk_B T_{c0}$, $mAk_B T$ or $mAk_B$. Only when $mAk_B T_{c0}$, $mAk_B T$ and $mAk_B$ are small, the series expansion method in this paper is meaningful.

From the analytical expression of the BEC critical temperature of equation (18), it can be seen that: when the GUP is considered, the critical temperatures $T_c$ of the ideal Bose gases trapped in the external power-law potentials in the form of equation (2) rise. Because the BEC is a kind of "cooperative" phenomenon of quantum correlation between particles. The GUP itself is an added quantum effect, which is similar to the addition of a quantum "correlation" between particles that enhances the "cooperation" phenomenon of the BEC and leads to an increase in $T_c$.

In order to further analyze the influence of the GUP on the thermodynamic properties of the ideal Bose gas at low temperature and the precise numerical relations, we take the ideal Bose gas trapped in 3D harmonic potential as an example. At this point, we know from equation (10) $X = 1/2 + 1/2 + 1/2 = 3/2$, $\eta = 3$. When $T < T_c$, the internal energy, heat capacity, entropy, critical temperature and excited states particle number are, respectively

$$U = 2.7 Nk_B T \left(\frac{T}{T_c}\right)^3 (1 + 8.1 mAk_B T_c - 11.496 mAk_B T + \cdots), \qquad (30)$$

$$C = 2.701 N k_B \left(\frac{T}{T_c}\right)^3 \left(4 + 32.414 m A k_B T_c - 57.483 m A k_B T + \cdots\right), \tag{31}$$

$$S = 0.900 N k_B \left(\frac{T}{T_c}\right)^3 \left(4 + 32.414 m A k_B T_c - 43.113 m A k_B T + \cdots\right), \tag{32}$$

$$T_c = T_{c0} \left[1 + 2.7 m A k_B T_{c0} - 3.993 \left(m A k_B T_{c0}\right)^2 + \cdots\right], \tag{33}$$

$$N_e = N \left(\frac{T}{T_c}\right)^3 \left(1 + 8.104 m A k_B T_c - 8.104 m A k_B T + \cdots\right). \tag{34}$$

When $A=0$ and $T<T_c$, equations (30)-(34) return to the expressions of the thermodynamic quantities of the ideal Bose gas in the 3D harmonic potential without considering the GUP.

$$U = 2.7 N k_B T \left(\frac{T}{T_c}\right)^3, \tag{35}$$

$$C = 10.8 N k_B \left(\frac{T}{T_c}\right)^3, \tag{36}$$

$$S = 3.600 N k_B \left(\frac{T}{T_c}\right)^3, \tag{37}$$

$$T_c = T_{c0} = \frac{1}{k_B} \left[\frac{N h^3}{\zeta(3)(2\pi m)^{3/2}} \prod_{i=1}^{3} \frac{\varepsilon^{1/t_i}}{g(2L_i)\Gamma(1/t_i + 1)}\right]^{1/3}, \tag{38}$$

$$N_e = N \left(\frac{T}{T_c}\right)^3. \tag{39}$$

When $T<T_c$, the amendments of the GUP to the internal energy, heat capacity, entropy, critical temperature and excited states particle number of idea Bose gas are, respectively

$$\Delta U = 2.7 N k_B T \left(\frac{T}{T_c}\right)^3 \left(8.1 m A k_B T_c - 11.496 m A k_B T + \cdots\right), \tag{40}$$

$$\Delta C = 2.701 N k_B \left(\frac{T}{T_c}\right)^3 \left(32.414 m A k_B T_c - 57.483 m A k_B T + \cdots\right), \tag{41}$$

$$\Delta S = 0.900 N k_B \left(\frac{T}{T_c}\right)^3 \left(32.414 m A k_B T_c - 43.113 m A k_B T + \cdots\right), \tag{42}$$

$$\Delta T = T_c - T_{c0} = T_{c0}\left[2.7mAk_BT_{c0} - 3.993(mAk_BT_{c0})^2 + \cdots\right], \tag{43}$$

$$\Delta N_e = N\left(\frac{T}{T_c}\right)^3 (8.104mAk_BT_c - 8.104mAk_BT + \cdots). \tag{44}$$

In some Refs of the quantum theory of gravity [33-35], the range of minimum observable length is given as $\Delta x_{min} = 10^{-17} \sim 10^{-16}$ m. In this paper, we let $\Delta x_{min} \approx 10^{-16}$ m. According to equation (5), the positive parameter $A$ can be regarded as $10^{36}$ (m/JS)$^2$. Note that the order of magnitudes of particle mass $m$, Boltzmann constant $k_B$ and critical temperature $T_{c0}$, and low temperature preconditions, such as, $mAk_BT_{c0}$, $mAk_BT$ and $mAk_B$ are usually small quantities. The order of magnitude of the first order correction term of the GUP is much larger than that of the second order correction term, so only the first order correction term of the GUP is considered.

As follows, we take two examples to do numerical calculation.

One is rubidium atom Bose gas. We take the particle density of ideal gas in the standard state as the average density of rubidium atoms ideal Bose gas, namely Loschmidt constant $n = N/V = 2.687 \times 10^{25}$ m$^{-3}$. For simplicity, we let $\varepsilon = \varepsilon_1 = \varepsilon_2 = \varepsilon_3 = 1$ J, and this is a very strong external potential. The mass of the rubidium atom $m$ equals to $1.445 \times 10^{-25}$ kg, and the spin degeneracy $g$ equals to 1. (See figures 1-7 and table 1.)

The other is sodium atoms Bose gas that originally verified the BEC [2]. The experimental data are: $n = 1.5 \times 10^{20}$ m$^{-3}$, $N = 7 \times 10^5$, $\omega_x = 745$ Hz, $\omega_y = 235$ Hz, $\omega_z = 410$ Hz, $m = 3.819 \times 10^{-26}$ kg, $g = 1$. Using $\varepsilon_i/L_i^{t_i} = m\omega_i^2/2$, we can estimate $\varepsilon_i \approx 10^{-31} \sim 10^{-30}$ J. (See figures 8-10 and table 2.)

Figure 1 shows that in 3D harmonic potential the numerical calculation results of the relations between the internal energy of ideal rubidium atomic gas with temperature, and the GUP's amendment to the internal energy (embedded figure) with temperature, when the GUP is, and is not, considered and $T < T_c$.

According to the comparison of the two internal energy variation curves with and without the GUP with temperature in figure 1, it can be known that: In the temperature region where $T/T_c < 0.71$, the GUP increases the internal energy at the same temperature. In the temperature region of $T/T_c > 0.71$, the GUP reduces the internal energy at the same temperature.

As the temperature increases, the "gravitational effect", a characteristic of the quantum correlation of the boson system, increases, and the additional energy of the gravitational effect is negative, which is represented by a decrease in internal energy. Because the GUP is derived from "quantum effect of gravity". From the embedded diagram in figure 1, it can be seen the amendment of the GUP of internal energy can reach an order of magnitude $10^{-1}$, and the GUP's influence is already large at this time.

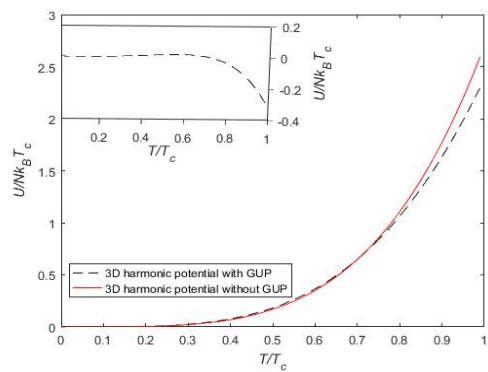

Figure 1. Changes of the numerical relationship between the internal energy and amendment to the internal energy (embedded graph) of the rubidium atoms ideal gas in a 3D harmonic potential with temperature when the particle density is Loschmidt constant, $\varepsilon=\varepsilon_1=\varepsilon_2=\varepsilon_3=1\mathrm{J}$ and $T<T_c$.

Figure 2 shows the numerical relationship of heat capacity and the GUP's amendment to the heat capacity with temperature, when the GUP is, and is not, considered and $T< T_c$. According to figure 2, the amendment of the GUP of the heat capacity can reach an order of magnitude $10^{-1}$, and the GUP's influence is already large at this time. In the temperature region where $T/T_c < 0.57$, the GUP increases the heat capacity at the same temperature. In the temperature region of $T/T_c > 0.57$, the GUP reduces the heat capacity at the same temperature.

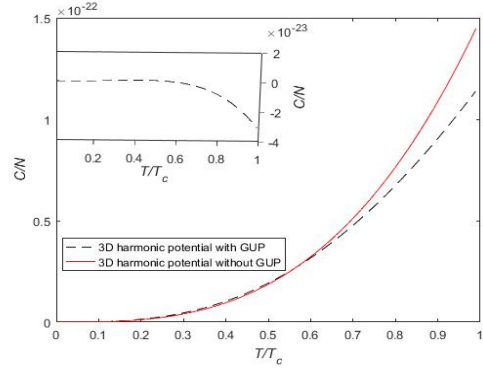

Figure 2. Changes of the numerical relationship between the heat capacity and amendment to the heat capacity (embedded graph) of the rubidium atoms ideal gas in a 3D harmonic potential with temperature when the particle density is Loschmidt constant, $\varepsilon=\varepsilon_1=\varepsilon_2=\varepsilon_3=1$J and $T<T_c$.

Figure 3 shows the numerical relationship of entropy and the GUP's amendment to the entropy with temperature, when the GUP is, and is not, considered and $T<T_c$. According to figure 3, the amendment of the GUP of the entropy can reach an order of magnitude $10^{-1}$. In the temperature region where $T/T_c < 0.73$, the GUP increases the entropy at the same temperature. In the temperature region of $T/T_c > 0.73$, the GUP reduces the entropy at the same temperature.

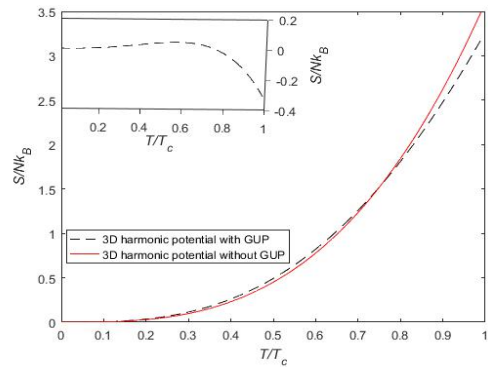

Figure 3. Changes of the numerical relationship between the entropy and amendment to the entropy (embedded graph) of the rubidium atoms ideal gas in a 3D harmonic potential with temperature when the particle density is Loschmidt constant, $\varepsilon=\varepsilon_1=\varepsilon_2=\varepsilon_3=1$J and $T<T_c$.

Figures 4, 5 and 6 are the comparison of the numerical relationship between the influence of the GUP on the internal energy, heat capacity and entropy with the change of temperature in different external potentials when $T<T_c$, respectively.

It can be seen from figures 4, 5 and 6 that with the increase of $X=\sum_{i=1}^{3}1/t_i$

($X$=0 is a free particle system, $X$=3/2 is a system in 3D harmonic potential) and temperature, the amendments of the internal energy, heat capacity and entropy of ideal rubidium atomic gas by the GUP decreases first and then increases near the turning point, and increases significantly after passing the turning point, that is, the external potential has a great influence on the amendments of the GUP. In order to further clearly display this characteristic, table 1 gives the specific values of the thermodynamic quantities of the ideal rubidium atomic gases in a container with a fixed volume and in a 3D harmonic potential when $\varepsilon=\varepsilon_1=\varepsilon_2=\varepsilon_3=1\mathrm{J}$, $T=T_c$, the GUP is considered and not considered, and when the density of rubidium atom is Loschmidt constant.

Comparing these values, it is known that the amendments of the GUP to the thermodynamic quantities of idea rubidium atom gas in the 3D harmonic potential increase by 13 to 26 orders of magnitude than the amendments of free ideal rubidium atom gas in the fixed container.

After further numerical analysis, we found that when $\varepsilon \geq 10^{-26}\mathrm{J}$, the critical temperature and internal energy of rubidium atoms ideal Bose gas with particle density of Loschmidt constant increased with the increase of $X$ at $T=T_c$; When $\varepsilon \leq 10^{-28}\mathrm{J}$, they decrease as $X$ increases at $T=T_c$. See figure 7. However, the heat capacity and entropy have no such characteristics.

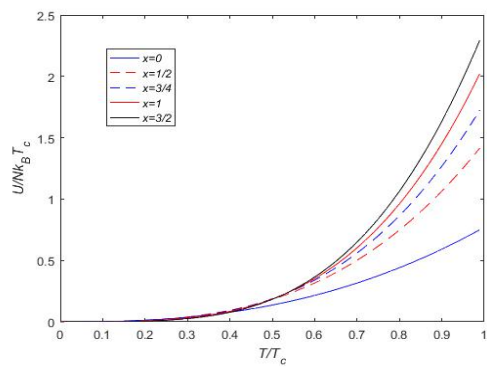

Figure 4. Comparison of the relation of the internal energy of the rubidium atoms ideal gas in different external potentials considering the GUP with temperature when the particle density is Loschmidt constant, $\varepsilon=\varepsilon_1=\varepsilon_2=\varepsilon_3=1\mathrm{J}$ and $T \leq T_c$.

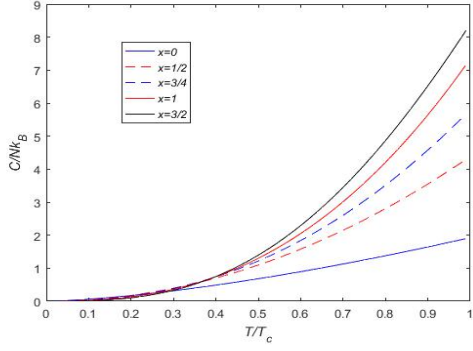

Figure 5. Comparison of the relation of the heat capacity of the rubidium atoms ideal gas in different external potentials considering the GUP with temperature when the particle density is Loschmidt constant, $\varepsilon=\varepsilon_1=\varepsilon_2=\varepsilon_3=1$J and $T \leq T_c$.

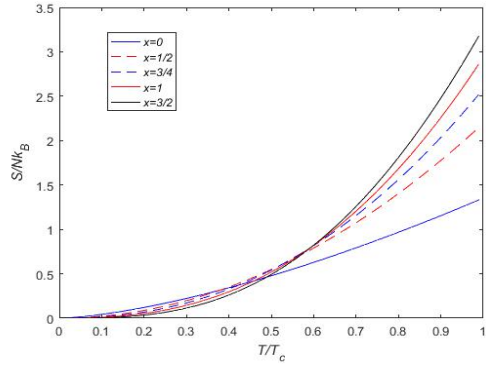

Figure 6. Comparison of the relation of the entropy of the rubidium atoms ideal gas in different external potentials considering the GUP with temperature when the particle density is Loschmidt constant, $\varepsilon=\varepsilon_1=\varepsilon_2=\varepsilon_3=1$J and $T \leq T_c$.

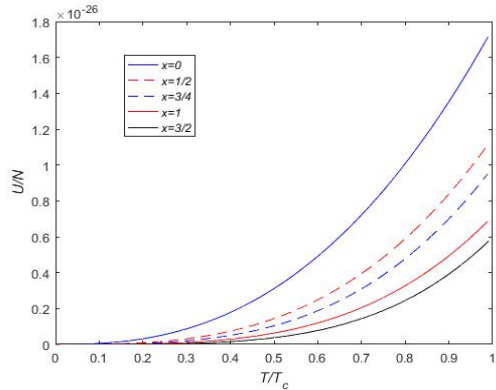

Figure 7. Comparison of the relation of the internal energy of the rubidium atoms ideal gas in different external potentials considering the GUP with temperature when the particle density is Loschmidt constant, $\varepsilon=\varepsilon_1=\varepsilon_2=\varepsilon_3=10^{-28}$J and $T \leq T_c$.

Table 1. Comparison of thermodynamic variables and their amendments of rubidium atom system between the system in the harmonic potential and the free particle system when the GUP is and not is, considered with the standard state.

| $N/V=2.687\times10^{25}\text{m}^{-3}$ $\varepsilon=1(\text{J})$ $T=T_c$ | Thermodynamic quantity of rubidium atom system without the GUP. | | The first order amendment term of the GUP to thermodynamic quantity of rubidium atom system. | |
|---|---|---|---|---|
| External potential | Free particle system. ($X=0$) | System in 3D harmonic potential. ($X=3/2$) | Free particle system.($X=0$) | System in 3D harmonic potential . ($X=3/2$) |
| Internal energy (J/particle) | $1.840\times10^{-26}$ | $6.641\times10^{-13}$ | $-4.642\times10^{-40}$ | $-7.556\times10^{-14}$ |
| Heat capacity (J/K·particle) | $2.658\times10^{-23}$ | $1.492\times10^{-22}$ | $-1.144\times10^{-36}$ | $-3.245\times10^{-23}$ |
| Entropy (J/K·particle) | $1.868\times10^{-23}$ | $4.927\times10^{-23}$ | $-3.679\times10^{-37}$ | $-4.615\times10^{-24}$ |

Because in the initial experiments to verify the BEC of the sodium atom gas, the atomic density ($n \approx 10^{20}\text{m}^{-3}$) is relatively low and the external field strength ($\varepsilon \approx 10^{-31}$-$10^{-30}$J) is relatively small, the relative amendment terms of the GUP are only $10^{-18}$-$10^{-17}$, and the effect of the GUP can be completely ignored. The numerical relationship of the internal energy, heat capacity and entropy and the GUP's amendment to these variables with temperature are similar to figures 1, 2 and 3, when the GUP is, and is not, considered and $T<T_c$.

Figures 8, 9 and 10 are the comparison of the numerical relationship between the influence of the GUP on the internal energy, heat capacity and entropy of the sodium atoms ideal gas with the change of temperature in different external potentials when $T<T_c$, respectively.

By comparing the numerical calculation results in table.2, we found that:

1）In the experimental scenario of the BEC of the sodium atoms gas, the atomic density and the strength of the external field, the influence of the external field and the GUP on the thermodynamic properties of the system is far less than that of the above theoretical scenario of rubidium atoms gas.

2）When $\varepsilon \geq 10^{-26}$ J, the critical temperature and internal energy of sodium atoms gas increased with the increase of $X$ at $T=T_c$; When $\varepsilon \leq 10^{-33}$ J, they decrease as $X$ increases at $T=T_c$. However, the heat capacity and entropy have also no such characteristics.

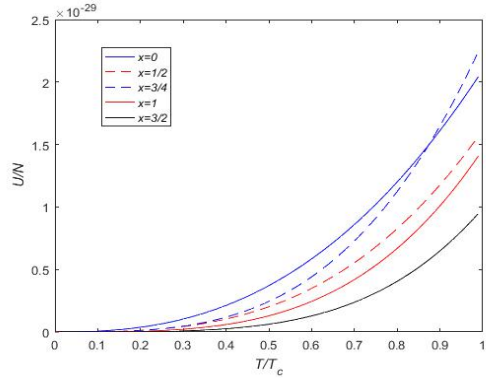

Figure 8. Comparison of the relation of the internal energy of the sodium atoms ideal gas in different external potentials considering the GUP with temperature when

$n = 1.5 \times 10^{20}$ m$^{-3}$, $\varepsilon$=9.211×10$^{-31}$ J and $T \leq T_c$.

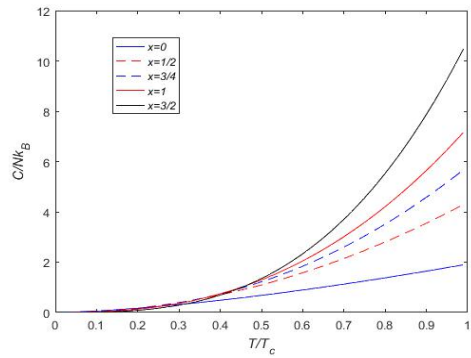

Figure 9. Comparison of the relation of the heat capacity of the sodium atoms ideal gas in different external potentials considering the GUP with temperature when

$n = 1.5 \times 10^{20}$ m$^{-3}$, $\varepsilon$=9.211×10$^{-31}$ J and $T \leq T_c$.

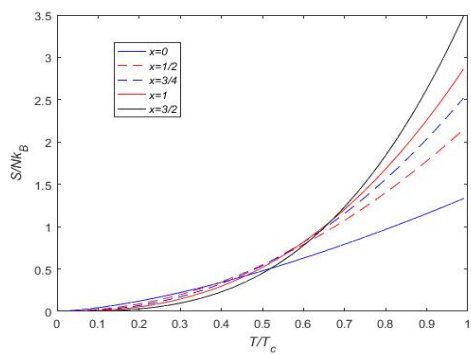

Figure 10. Comparison of the relation of the entropy of the sodium atoms ideal gas in different external potentials considering the GUP with temperature when $n = 1.5 \times 10^{20} \text{m}^{-3}, \varepsilon = 9.211 \times 10^{-31} \text{J}$ and $T \leq T_c$.

Table 2. Comparison of thermodynamic variables and their amendments of sodium atom system between the system in the harmonic potential and the free particle system when the GUP is and not is, considered with the experimental data of verifying the BEC for the first time.

| $n=1.5\times 10^{20}\text{m}^{-3}$ $\varepsilon \approx 9.211\times 10^{-31}(\text{J})$ $N=7\times 10^5$. $T=T_c$ | Thermodynamic quantity of sodium atom system without the GUP. | | The first order amendment term of the GUP to thermodynamic quantity of sodium atom system. | |
|---|---|---|---|---|
| External potential | Free particle system. ($X=0$) | System in 3D harmonic potential. ($X=3/2$) | Free particle system. ($X=0$) | System in 3D harmonic potential. ($X=3/2$) |
| Internal energy (J/particle) | $2.097 \times 10^{-29}$ | $9.885 \times 10^{-30}$ | $-1.740 \times 10^{-46}$ | $-4.636 \times 10^{-48}$ |
| Heat capacity (J/K·particle) | $2.658 \times 10^{-23}$ | $1.492 \times 10^{-22}$ | $-3.599 \times 10^{-40}$ | $-1.307 \times 10^{-40}$ |
| Entropy (J/K·particle) | $1.868 \times 10^{-23}$ | $4.972 \times 10^{-23}$ | $-1.158 \times 10^{-40}$ | $-1.859 \times 10^{-41}$ |

## 4. Conclusion

In this paper, the generalized uncertainty principle (GUP) or the quantum effect of gravity is considered. On the premise that the amendments of the GUP are small, the critical temperature $T_c$ and the modification of $T_c$ by the GUP for the Bose Einstein condensate (BEC) of ideal Bose gases in different external power law potentials given in equation (2) are calculated analytically [see equations (18) and (19)]. Under the conditions of $T > T_c$ and $T < T_c$, the analytical expressions of the internal energy, heat capacity, entropy, ground state and excited state particle numbers are given respectively[see equations (20)-(22), (23)-(29)]. Taking the 3D harmonic potential as an example, when the particle densities are Loschmidt constant ($n \approx 10^{25}\text{m}^{-3}$) and the initial experimental data to verify the BEC ($n \approx 10^{20}\text{m}^{-3}$), the quantities $\varepsilon = \varepsilon_1 = \varepsilon_2 = \varepsilon_3$ reflecting the external potential intensity are 1J and

9.211×10⁻³¹J respectively and the temperature is lower than the critical temperature $T_c$, the relation of the internal energy, heat capacity, entropy and the GUP's amendments of rubidium and sodium atoms ideal Bose gases with temperature are calculated and analyzed numerically, and the conclusions are as follows:

1) The GUP raises the temperature of the BEC of the ideal Bose gas, whether it is in a container with a fixed volume or in an external potential. Because the "minimum observable length" implied by the GUP leads to the enhancement of this quantum "cooperation" phenomenon between the BEC particles.

2) When $T<T_c$ and $T$ is slightly greater than 0K, the amendments of the GUP to the internal energy, heat capacity and entropy are positive. When the temperature rises to a certain value, these amendments become negative, that is, thermodynamic variables such as the internal energy, heat capacity and entropy increase first and then decrease under the influence of the GUP. We think that " increase first " is the direct expression of the state density decrease. The "then decrease" is because the GUP originates from the "quantum effect of gravity". The characteristic of the quantum correlation of the multi-particle boson system is the "gravitational effect". The GUP enhances the quantum correlation of the boson system, which is manifested as the decrease of the internal energy.

3) External potentials can make the GUP's corrections huge. Under the condition that $\varepsilon=\varepsilon_1=\varepsilon_2=\varepsilon_3=1J$ and $n=2.687\times10^{25}m^{-3}$ are taken, due to the correlation between the external potential and the GUP, the GUP's amendments of the internal energy, heat capacity and entropy of the rubidium atom gas decreases first and then increases with the increase of $X$ ($X\equiv\Sigma_i 1/t_i$ is the sum of reciprocal of the exponents $t_i$ of the power function) and temperature. When $T=T_c$, compared with the free particle system with $X=0$, the amendments of the GUP to the internal energy, heat capacity and entropy of the system in the 3D harmonic potential with $X=3/2$ increases by 26, 13 and 13 orders of magnitude respectively.

4) The $\varepsilon$ has a transition region ($\varepsilon_a, \varepsilon_b$). When $\varepsilon>\varepsilon_b$, the amendments of the GUP to the internal energy and critical temperature increased with the increase of $X$ at $T=T_c$; When $\varepsilon<\varepsilon_a$, they decrease as $X$ increases at $T=T_c$. However, the heat capacity and entropy have no such characteristics.

5) When $T<T_c$ and there is no external potential, the amendments of the GUP is usually small. For example, the relative amendments of the thermodynamic variables of free rubidium atoms ideal gas in a fixed volume with an average particle density of Loschmidt constant are only $10^{-14}$ to $10^{-13}$ orders of magnitude, and the influence of the GUP can be completely ignored. However, under the condition of the same particle density and temperature, the relative amendments can reach the order of magnitude of $10^{-1}$ when the external potential intensity $\varepsilon=\varepsilon_1=\varepsilon_2=\varepsilon_3$ is 1J in the 3D harmonic potential, and the GUP's amendments should be considered.

6) Under the experimental situation of particle density $n \approx 10^{20} m^{-3}$ and the external potential intensity $\varepsilon \approx 10^{-31}$J, the relative amendments of the GUP to the thermodynamic variables are about $10^{-18}$-$10^{-17}$. At this point, the effect of the GUP can be completely ignored.

7) The GUP may dominate the properties of a system rather than act as a simple amendment, which cannot be accurately analyzed by the series expansion method proposed in this paper and needs to be confirmed by a more accurate calculation method.

**References:**

[1] Anderson M H, Ensher J R, Matthews M R, Wieman C E, Cornell E A 1995 *Science* **269** 198

[2] Davis K B, Mewes M-O, Andrews M R, van Druten N J, Durfee D S, Kurn D M, Ketterle W 1995 *Phys. Rev. Lett.* **75** 3969

[3] Bradley C C, Sackett C A, Tollett J J, Hulet R G 1995 *Phys. Rev. Lett.* **75** 1687

[4] Dalfovo F, Giorgini S, Pitaevskii L P, Stringari S 1999 *Rev. Mod. Phys.* **71** 463

[5] Leggett A J 2001 *Rev. Mod. Phys.* **73** 307

[6] Pethick C J, Smith H 2002 *Bose-Einstein Condensation in Dilute Gases* (Cambridge: Cambridge University Press)

[7] Bongs K, Sengstock K 2004 *Rep. Prog. Phys.* **67** 907

[8] Anglin J R, Ketterle W 2002 *Nature* **416** 211

[9] Rolston S L, Phillips W D 2002 *Nature* **416** 219

[10] Regal C A, Greiner M, Jin D S 2004 *Phys. Rev. Lett.* **92** 040403

[11] Jochim S, Bartenstein M, Altmryer A, Hendl G, Riedel S, Chin C, Denschlag J H, Grimm R 2003 *Science* **302** 2101

[12] O'Hara K M, Hemmer S L, Gehm M E, Granade S R, Thomas J E 2002 *Science* **298** 2179

[13] Vanderlei Bagnato, David E, Pritchard, Daniel Kleppner 1987 *Phys. Rev. A* **35** 4354

[14] Shi H L, Zheng W M 1997 *Phys. Rev. A* **56** 1046

[15] Heqiu Li, Qiujiang Guo, Ji Jiang, and D. C. Johnston 2015 *Phys. Rev. E* **92** 062109

[16] Su G ZH, Chen J C 2011 *Journal of Xiamen University* (Natural Science) **50** 217.

[17] Lixuan Chen, Zijun Yan, Mingzhe Li and Chuanhong Chen 1998 *J. Phys. A: Math. Gen*. **31** 8289.

[18] Chou T T, Yang C N, Yu L H 1996 *Phys. Rev. A* **53** 4257

[19] Kempf A, Mangano G, Mann R B 1995 *Phys. Rev. D* **52** 1108

[20] Kempf A "*On the structure of space-time at the Planck scale*" *Proc. 36th Course Int.* 2000; (arXiv:hep-th/9810215 or arXiv:hep-th/9810215v1)

[21] Amati D, Ciafaloni M, Veneziano G 1989 *Phys. Lett. B* **216** 41

[22] Garay L J 1995 *Int. J. Mod. Phys*. *A* **10** 145

[23] Scardigli F 1999 *Phys. Lett. B* **452** 39

[24] Chang L N, Minic D, Takeuchi T, Okamura N 2002 *Phys. Rev. D* **65** 125028

[25] Fityo V T 2008 *Phys. Lett. A* **372** 5872

[26] Chang L N, Minic D, Okamura N, Takeuchi T 2002 *Phys. Rev. D* **65** 125027

[27] Vakili B, Gorji M A 2012 *J. Stat. Mech*. P10013

[28] Li X 2002 *Phys. Lett. B* **540** 9

[29] Brout R, Gabriel Cl, Lub M, Spindel Ph 1999 *Phys. Rev. D* **59** 044005

[30] Zhao H H, Li G L, Zhang L C 2012 *Phys. Lett. A* **376** 2348

[31] Li H L, Wang J J, Yang B, Wang Y N, Shen H J 2015 *Acta Phys. Sin.* **64** 08502

[32] Li H L, Ren J X, Wang W W, Yang B, Shen H J 2018 *J. Stat. Mech*. 023106

[33] Zhao R, Zhang L C, Li H F 2009 *Acta Phys. Sin*. **58** 2193.

[34] Brau F 1999 *J. Phys. A: Math. Gen*. **32** 7691

[35] Stetsko M M, Tkachuk V M 2006 *Phys. Rev. A* **74** 012101